\documentclass[reprint,amsmath,amssymb,aps,prl,superscriptaddress,floatfix]{revtex4-1}
\usepackage{graphicx}
\usepackage{dcolumn}
\usepackage{bm}

\begin{document}


\title{Filters for High Rate Pulse Processing}

\author{B.K.~Alpert}
\author{R.D.~Horansky}
\author{D.A.~Bennett}
\author{W.B.~Doriese}
\author{J.W.~Fowler}
\affiliation{National Institute of Standards and Technology,
 Boulder, Colorado 80305}
\author{A.S.~Hoover}
\author{M.W.~Rabin}
\affiliation{Los Alamos National Laboratory,
  Los Alamos, New Mexico 87545}
\author{J.N.~Ullom}
\affiliation{National Institute of Standards and Technology,
 Boulder, Colorado 80305}

\begin{abstract}
  We introduce a filter-construction method for pulse processing that
  differs in two respects from that in standard optimal filtering, in
  which the average pulse shape and noise-power spectral density are
  combined to create a convolution filter for estimating pulse
  heights.  First, the proposed filters are computed in the time
  domain, to avoid periodicity artifacts of the discrete Fourier
  transform, and second, orthogonality constraints are imposed on the
  filters, to reduce the filtering procedure's sensitivity to unknown
  baseline height and pulse tails.  We analyze the proposed filters,
  predicting energy resolution under several scenarios, and apply the
  filters to high-rate pulse data from gamma-rays measured by a
  transition-edge-sensor microcalorimeter.
\begin{description}
\item[Keywords] baseline insensitivity, energy resolution, optimal
  filtering, pulse pile-up, pulse tail insensitivity
\item[PACS numbers] 07.20.Mc, 07.05.Kf, 84.30.Sk.
\item[U.S. government publication] Not subject to copyright.
\end{description}
\end{abstract}

\maketitle

The extraction of physical quantities from noisy data streams is
ubiquitous in the physical sciences.  Examples include the
determination of photon and particle energies or incidence times in
nuclear and particle physics.  Raw data records are invariably
filtered to extract the quantity of interest with the highest
signal-to-noise, and extensive effort has gone into filter development.
One important example is so-called ``optimal filtering,'' for isolated
pulses with amplitude proportional to photon energy.  Filters
constructed from the average pulse shape and the noise power spectral
density are convolved with pulse records to estimate pulse amplitudes
\cite{Szymkowiak93}.  This filter is widely used in X-ray astrophysics
\cite{Eckart12} and direct searches for weakly interacting dark matter
\cite{Ahmed11}.

Here, we propose and demonstrate a novel method for optimal-filter
construction for pulse processing.  The previous optimal filter is
shown to be an example of a much larger class of filters that have new
and useful properties.  For example, optimal filters can be
constructed that are orthogonal to exponential tails of prior pulses,
prompted by the need to cope with high-rate pulse data.  Many
applications of high-resolution photon spectroscopy require very large
photon counts for accurate characterization of an absorption or
emission spectrum across a broad energy band.  For example, isotopic
analysis of nuclear materials for treaty
verification requires approximately $10^9$ photons in a spectrum
between 60 keV and 260 keV to achieve
uncertainty of $10^{-3}$ \cite{Jethava09}.  Low-temperature detectors
can reach this goal, within limited collection periods, only through
large arrays of elements operating at high photon count rates per
element.  Consequently, operation at high count rates is an active
topic of research \cite{HuiTan08, HuiTan09, HuiTan11, Alpert12}.

The new framework departs from prior algorithms in two respects: (1)
noise autocovariance is used in place of its mathematical dual, the
noise power spectral density, to avoid the discrete Fourier transform
(DFT) and enable the construction to be entirely in the time domain;
and (2) the filter optimization is subject to explicit constraints
beyond maximization of signal-to-noise ratio for isolated pulses,
including for the filter length, orthogonality to constants, and
orthogonality to exponentials of one or more decay rates.  (The method
is related to constrained optimization in some other contexts.  For
example, a similar approach has recently been developed for designing
matched filters for wavefront sensing \cite{Gilles08}.)  Orthogonality
to exponentials can reduce or eliminate sensitivity to tails of prior
pulses.  With these additional constraints imposed, the filters suffer
some loss of sensitivity for isolated pulses compared to filters
optimized for that case, but they compensate by retaining resolution
with piled-up pulses and by avoiding DFT artifacts, including
artificial periodicity.

\paragraph*{Processing procedure.}
Estimation of pulse amplitudes under standard filtering
\cite{Szymkowiak93, Moseley88} is optimized for isolated pulses.  Each
pulse is convolved with a filter and the maximum of the convolution
(or a smoothed maximum as provided by a quadratic polynomial fit to
several values near the maximum) provides an estimate of the pulse
amplitude.  The filter, in principle, is constructed to minimize the
variance of this estimate, given a known pulse shape and known noise
power spectrum.

\paragraph*{Continuous time model.}
We assume a signal $f$ consists of a pair of pulses sitting on a baseline
\[
f(t)=a_0 s(t-t_0)+a_1 s(t-t_1)+b,\label{signal}
\]
where $s$ is the pulse shape, $a_0$ and $a_1$ are the pulse
amplitudes, $t_0$ and $t_1$ are the pulse arrival times, with
$t_0<t_1$, and $b$ is the baseline.  A noisy signal
consists of signal plus noise, $m(t)=f(t)+\eta(t),$ where the noise
$\eta$ is assumed to be a realization of a stationary stochastic
process with a mean of zero and autocovariance
\[
R_\eta(\tau)=\int_{-\infty}^\infty \eta(t)\eta(t+\tau)dt.
\]

\paragraph*{Discrete time model.}
The measurement apparatus obtains an approximation $m_i$ of $m(i\Delta)$
for $i$ an integer, where $\Delta$ is the sample time spacing,
as a convolution of $m$ with a response function
\[
m_i=\int_{-\infty}^\infty m(i\Delta-t)w(t)dt,
\]
where $w$ is an approximate $\delta$-function centered
at the origin with unit integral.  We define $f_i,$ $s_i,$ and $\eta_i$
analogously.  Our measurement model is then
\begin{align}
m_i&=f_i+\eta_i\nonumber\\
&=a_0 s_{i-i_0}+a_1 s_{i-i_1}+b+\eta_i.\label{discrete}
\end{align}
In this approximate model, arrival times $t_0=i_0\Delta,\;t_1=i_1\Delta$
are assumed
aligned with the samples, and known, to avoid interpolation issues. The
pulse shape $s=(s_0,\ldots,s_n,\ldots)^t$ is approximated by averaging
many pulses to obtain the estimate
$\hat{s}=(\hat{s}_0,\ldots,\hat{s}_n,\ldots)^t$, normalized so
$\max\hat{s}=1$, and the noise autocovariance
$r=(r_0,\ldots,r_n,\ldots)^t$, given by the expectation
\begin{equation}
r_k=\mathbb{E}\left[\eta_i\eta_{i+k}\right]-\mathbb{E}\left[\eta_i\right]^2
=\mathbb{E}\left[\eta_i\eta_{i+k}\right]\label{r},
\end{equation}
is approximated by averaging products of pulse-free samples of the sensor
output to obtain the estimate $\hat{r}=(\hat{r}_0,\ldots,\hat{r}_n,\ldots)^t$.

\paragraph*{Amplitude estimation.}
The standard procedure assumes $a_0=0$, computes the discrete
convolution
\[
(q\star m)_i=\sum_{j=0}^{n-1}q_{j}m_{i-j}
\]
of a given filter $q=(q_0,\ldots,q_{n-1})^t$ with
$\ldots,m_{-1},m_0,m_1,\ldots,$ the discrete convolution of $q$ with
$\ldots,\hat{s}_{-1},\hat{s}_0,\hat{s}_1,\ldots,$ where $\hat{s}_i=0$
for $i<0,$ and estimates $a_1$ as the ratio of their maximums
\begin{equation}
\hat{a}_1=\frac{\max_i(q\star m)_i}{\max_i(q\star \hat{s})_i}.
\end{equation}

\paragraph*{Estimate mean and variance.}
We seek the mean and variance of the amplitude estimate $\hat{a}_1.$
We have
\begin{equation}
  \mathbb{E}\left[(q\star m)_i\right]
  =a_0\cdot (q\star s)_{i-i_0}+a_1\cdot (q\star s)_{i-i_1}+b\sum_{j=0}^{n-1}q_j.
\end{equation}
We define $\bar{\imath}$ so that $(q\star s)_{\bar{\imath}}=\max_i(q\star s)_i.$
Under assumptions of orthogonality to the prior tail and to constants,
\begin{equation}
(q\star s)_{\bar{\imath}+i_1-i_0}=0=\sum_{j=0}^{n-1}q_j,\label{constraints}
\end{equation}
we have
\begin{align}
\mathbb{E}\left[\hat{a}_1\right]&=\frac{\mathbb{E}\left[\max_i(q\star m)_i
\right]}{\max_i(q\star \hat{s})_i}\nonumber\\
&\approx\frac{\max_i\mathbb{E}\left[(q\star m)_i\right]}{
\max_i(q\star \hat{s})_i}
\approx\frac{a_1\cdot (q\star s)_{\bar{\imath}}}{(q\star\hat{s})_{\bar{\imath}}}
\approx a_1,
\end{align}
where the approximations are equalities under somewhat restrictive
conditions.

Toward a variance estimate,
\begin{multline*}
  \mathbb{E}\left[m_{i-j}m_{i-k}\right]=
\left(a_0 s_{i-i_0-j}+a_1 s_{i-i_1-j}+b\right)\\
\times\left(a_0 s_{i-i_0-k}+a_1 s_{i-i_1-k}+b\right)+r_{j-k},
\end{multline*}
where $r_{j-k}$ is the noise autocovariance of (\ref{r}).  Now
\begin{align}
{\rm Var}\left[\hat{a}_1\right]&=\mathbb{E}\big[{\hat{a}_1}^{\;2}\big]
-\mathbb{E}\left[\hat{a}_1\right]^2\nonumber\\
&=\frac{\mathbb{E}\big[\max_i(q\star m)_i^{\;2}\big]
  -\mathbb{E}\left[\max_i(q\star m)_i\right]^2}{\max_i(q\star \hat{s})_i^{\;2}}
\nonumber
\end{align}
\begin{align}
\hspace{0.49in} &\approx\frac{\max_i\mathbb{E}\big[(q\star m)_i^{\;2}\big]
  -\max_i\mathbb{E}\left[(q\star m)_i\right]^2}{\max_i(q\star \hat{s})_i^{\;2}}
\nonumber\\
&=\frac{q^tRq}{[q^t\overline{s}]^2}\approx\frac{q^t\hat{R}q}{[q^t\overline{s}]^2}
\stackrel{\mathrm{def}}{=}\hat{\rm V}{\rm ar}\left[\hat{a}_1\right],
\end{align}
where the variance estimate $\hat{\rm V}{\rm ar}\big[\hat{a}_1\big]$
is defined to be the last expression, $\hat{R}$ is the $n\times n$
estimated covariance matrix with $\hat{R}_{jk}=\hat{r}_{j-k}=\hat{r}_{|j-k|},$ and
$\overline{s}=(\hat{s}_i,\hat{s}_{i-1},\ldots,\hat{s}_{i-n+1})^t$ is
the length $n$ segment from $\hat{s}$ with
$q^t\overline{s}=\max_i(q\star\hat{s})_i$.

\begin{figure}
\includegraphics[width=1.095\linewidth,keepaspectratio]{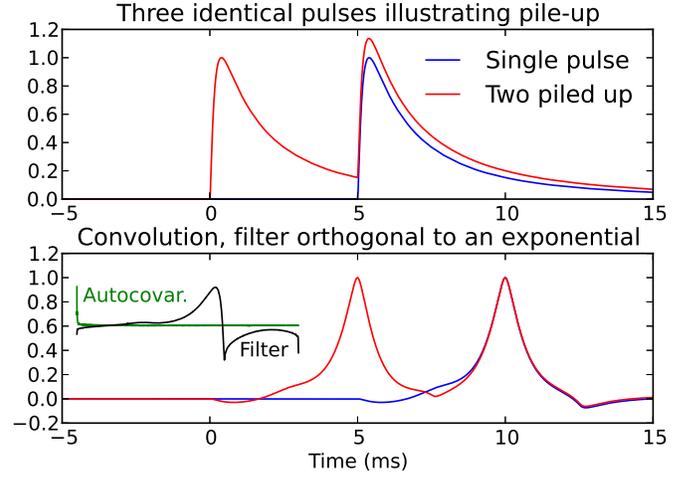}
\caption{\label{convol}(color) Two scenarios, one with pile-up, are shown
  (top). From the pulse shape and noise autocovariance, a filter
  orthogonal to an exponential of tail decay, $\tau=3.2$ ms, is
  computed (inset, separate vertical scales).  Convolution of the filter
  with the signal yields peaks of essentially constant height
  (bottom) and nearly eliminates pile-up dependence.}
\end{figure}

\begin{table}
  \caption{\label{summary}Dataset summary for four photon count rates.
    Sample spacing was 2.56 $\mu$s, full stream data were kept, and records
    were later formed, including for two lengths shown. Discards resulted
    from insufficient time between triggered pulses, signal drops below
    baseline, and rises in pre-trigger or post-peak attributed to
    (untriggered) nuisance pulses.}
\begin{tabular}{|l|r|r|r|r|}
\hline
Dataset & \multicolumn{1}{c|}{1} & \multicolumn{1}{c|}{2}
& \multicolumn{1}{c|}{3} & \multicolumn{1}{c|}{4} \\
\hline
Duration (s) & 4249.50 & 3096.93 & 3182.13 & 4583.17 \\
Pulses triggered & 5496 & 6581 & 17872 & 60267 \\
Rate (Hz)        & 1.29 & 2.13 &  5.62 & 13.15 \\
\hline
\multicolumn{5}{l}{Records: 10.24 ms with 5.12 ms pre-trigger}\\
\hline
Discards: & & & & \\
\hspace{0.2in}pulse starts ($>$1)& 143 & 201 & 1088 &  7610 \\
\hspace{0.2in}SQUID unlock       &  42 &  92 &  585 &  3648 \\
\hspace{0.2in}early peak         &  22 &  17 &   62 &   167 \\
\hspace{0.2in}pre-trigger rise   &   8 &  21 &   93 &   368 \\
\hspace{0.2in}post-peak rise     &   5 &  13 &   65 &   631 \\
\hline
97 keV (raw height) & 1095 & 1286 & 3213 & 9607 \\
\hline
\multicolumn{5}{l}{Records: 25.60 ms with 6.40 ms pre-trigger}\\
\hline
Discards: & & & & \\
\hspace{0.2in}pulse starts ($>$1)& 243 & 387 & 2463 & 17266 \\
\hspace{0.2in}SQUID unlock       &  40 &  90 &  537 &  2985 \\
\hspace{0.2in}early peak         &  19 &  17 &   58 &   138 \\
\hspace{0.2in}pre-trigger rise   &   8 &  23 &  105 &   606 \\
\hspace{0.2in}post-peak rise     &  19 &  36 &  223 &  1325 \\
\hline
97 keV (raw height) & 1067 & 1228 & 2938 & 7565 \\
\hline
\end{tabular}
\end{table}

\paragraph*{Filter optimization.}
This expression for the variance of the amplitude estimate
enables us to design filters that minimize the estimated variance.  
$\hat{\rm V}{\rm ar}\big[\hat{a}_1\big]$ is minimized at a stationary
point of the Lagrange function,
\[
\Lambda(q,\lambda)=q^t\hat{R}q-\lambda\left[q^t\overline{s}-1,\right]
\]
where $\lambda$ is a Lagrange multiplier
to ensure that the scale of
$q$ satisfies $q^t\overline{s}=1=\max\overline{s}.$ Setting
the partial derivatives of $\Lambda$ to zero and solving gives
\begin{equation}
q=\frac{\hat{R}^{-1}\overline{s}}{\overline{s}^t\hat{R}^{-1}\overline{s}},
\qquad\hat{\rm V}{\rm ar}\big[\hat{a}_1\big]=
\left[\overline{s}^t \hat{R}^{-1}\overline{s}\right]^{-1}.
\label{vanilla}
\end{equation}
This solution depends on the choice, made above tacitly, of the length $n$ of
the convolution filter $q$.

Extending beyond the prescription above, we optimize
subject to stipulated constraints.  Orthogonality to constants or
exponentials of particular decay rates can be imposed by revising the
Lagrange function.  For orthogonality to $k$ vectors
$V=\left[v_1\cdots v_k\right],$
\[
\Lambda(q,\lambda,\gamma)=q^t\hat{R}q-\lambda\left[q^t\overline{s}-
1\right]-q^tV\gamma,
\]
where $\gamma=(\gamma_1,\ldots,\gamma_k)^t$ are $k$ additional Lagrange
multipliers.  The solution is
\begin{equation}
q=\hat{R}^{-1}\overline{V}\left(\overline{V}^t\hat{R}^{-1}\overline{V}\right)^{-1}e_1,
\qquad\hat{\rm V}{\rm ar}\big[\hat{a}_1\big]=q^t\hat{R}q,
\end{equation}
where $\overline{V}=[\overline{s}\; v_1 \cdots v_k]$ and
$e_1=(1,0,\ldots,0)^t$ is of length $k+1.$

\begin{figure}
\includegraphics[width=1.0\linewidth,keepaspectratio]{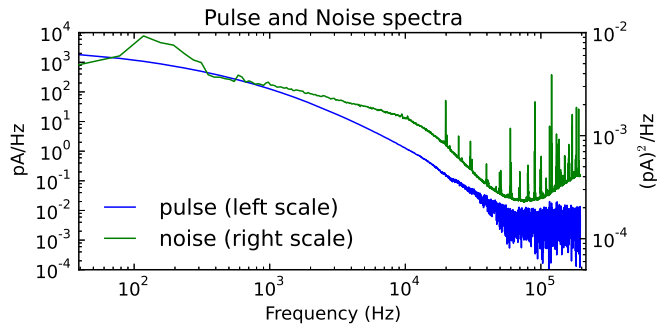}
\caption{\label{pulse_noise}(color) Pulse spectrum is the absolute
  value of the discrete Fourier transform (DFT) of the average of
  pulses, near 97.431 keV line, from the highest-rate dataset.  The
  noise spectrum is the average of the square of absolute value of
  DFT of pulse-free records of TES output. Records are 25.6 ms.}
\end{figure}
\begin{figure}
\includegraphics[width=1.09\linewidth,keepaspectratio]{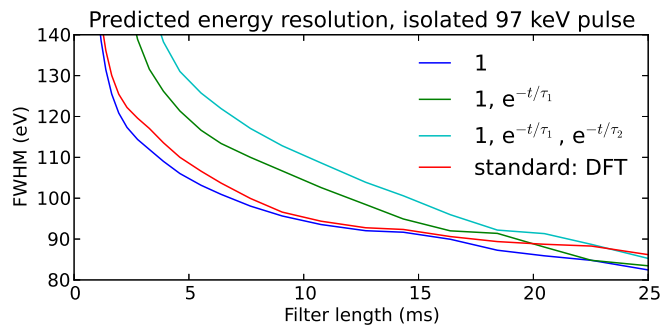}
\caption{\label{filtlen}(color) Predicted resolution on an isolated pulse
  of four filters is shown.  The filters, determined from average pulse
  shape and noise autocovariance (Fig.~\ref{pulse_noise}), include the
  standard DFT-computed filter with lowest frequency bin set to zero
  \cite{Doriese09}
  and proposed filters orthogonal to constants and zero, one, or two
  exponentials ($\tau_1=6.0$ ms, $\tau_2=1.5$ ms).}
\end{figure}
\begin{figure}
\includegraphics[width=1.07\linewidth,keepaspectratio]{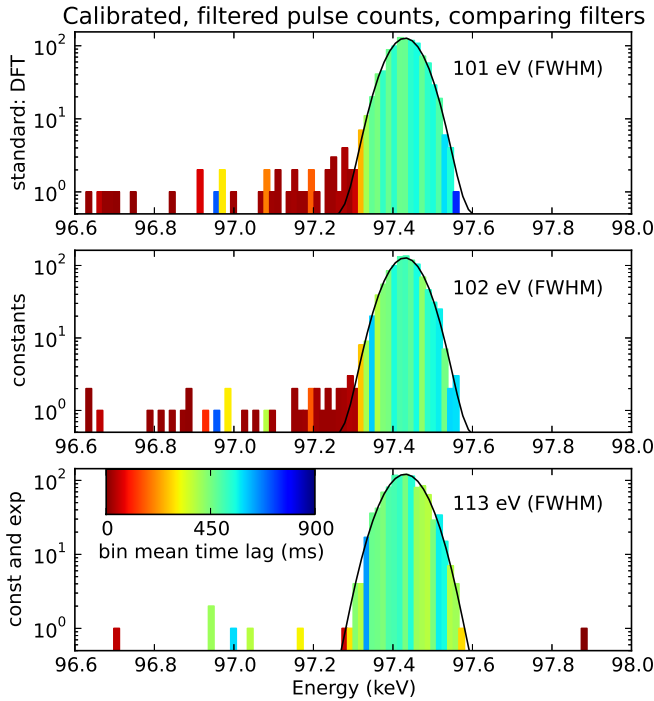}
\caption{\label{hist1}(color) Energy histograms near the 97.431 keV line,
  from the 2.13 Hz dataset with 10.24 ms record length, are shown for
  the standard filter, the proposed filter orthogonal to constants, and
  the proposed filter orthogonal to constants and exponentials
  ($\tau=6$ ms).  Color denotes the pulse arrival time lag since the
  previous pulse, averaged over the histogram bin, and illustrates
  that filtering errors, concentrated in heavily piled-up pulses, are
  nearly eliminated by the filter orthogonal to exponentials.}
\end{figure}

Orthogonality to exponentials of a particular decay time constant
enables filters to be less sensitive to tails of prior pulses.
Fig.~\ref{convol} illustrates the principle of these filters.
Avoidance of the DFT, with an increase in filter computation cost that is
very mild for filter lengths up to $n\approx 10^4,$ avoids false
assumptions of signal and noise periodicity and yields nonperiodic
filters.

\paragraph*{Experiment.}
Measurements were taken at NIST of photons from a $^{153}$Gd source
with a single transition-edge-sensor (TES) microcalorimeter \cite{Bennett12},
at varied
count rates (1.29, 2.13, 5.62, and 13.15 Hz), by placing the source at
four different distances from the detector.  Essentially all pulses
were filtered; no attempt was made to selectively discard pulses to
improve the energy resolution.  In extraction of pulse records from
the data streams, pulses were lost principally due to onset within
the prior pulse record and to occasional SQUID mode unlock.
Statistics for these measurements are summarized in Table~\ref{summary}.
The following analysis focuses on pulses near the 97.431 keV gamma-ray
emission line of $^{153}$Gd.

\begin{figure}
\includegraphics[width=1.07\linewidth,keepaspectratio]{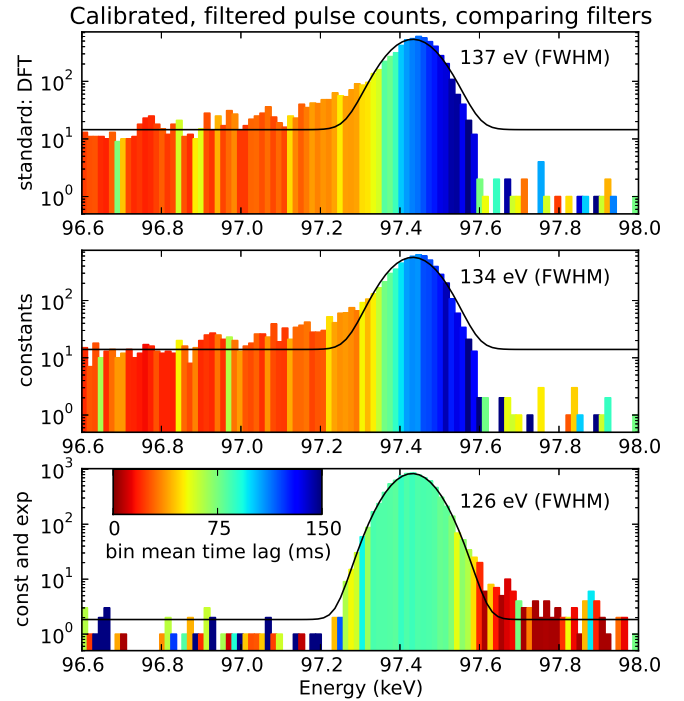}
\caption{\label{hist2}(color) Energy histograms as in Fig.~\ref{hist1},
  except from 13.15 Hz dataset.  At this higher rate, the errors of the
  first two filters are much more significant, as is the improvement
  offered by the third.}
\end{figure}

The noise spectrum and, for comparison, the spectrum of the average
pulse 
are plotted in Fig.~\ref{pulse_noise}.  The noise spectrum and the DFT
of the average pulse are used to compute the standard filter.  The
autocovariance and the average pulse (shown above in
Fig.~\ref{convol}) are used to compute the proposed filters and the
predicted energy resolution of each.  Fig.~\ref{filtlen} shows predicted
resolution versus filter length for the proposed filters and the
standard DFT-computed optimal filter, with the lowest frequency bin
set to zero to reduce sensitivity to baseline drift.  The standard
filter and the proposed filter orthogonal to constants would agree,
absent discretization and periodicity artifacts due to the DFT.  This
calculation is for isolated pulses; for piled-up pulses these two
filters suffer bias problems that are significantly reduced by the filters
orthogonal to exponentials.  The filter orthogonal to two
exponentials, however, due to the additional constraint, suffers
significant loss of sensitivity at short to moderate filter lengths,
and is not considered further here.

The performance of the other three filters is compared on measured
pulses, and histograms near the 97.431 keV line are plotted for two
different pulse rates in Fig.~\ref{hist1} and Fig.~\ref{hist2}.  Each
histogram was fit with a Gaussian plus a constant to determine the
energy resolution.  The histogram bins are colored based on the pulse
arrival-time lag from the previous pulse, averaged over the bin,
demonstrating that errors in processing are due mainly to closely
piled-up pulses and are significantly ameliorated by the proposed
filters orthogonal to exponentials.  This effect is pronounced at
the higher pulse rate, yielding much-enhanced peak height and reduced
leakage for the filter orthogonal to exponentials.

\begin{figure}
\includegraphics[width=1.08\linewidth,keepaspectratio]{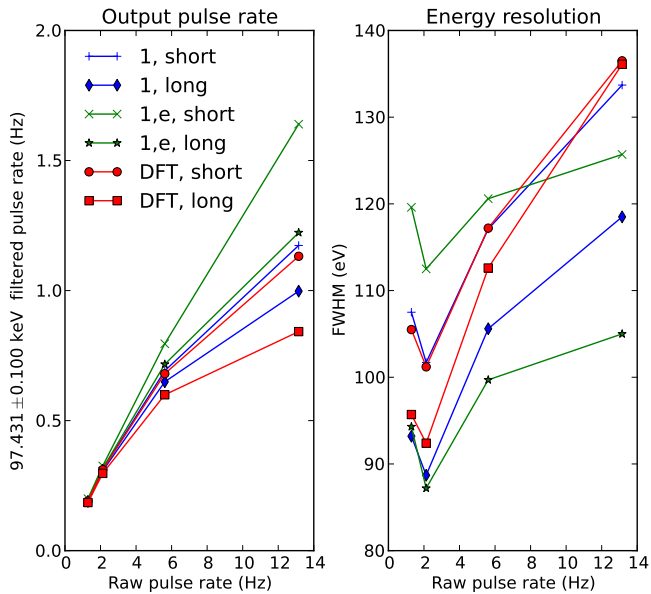}
\caption{\label{icr_ocr}(color) The output pulse rate and energy resolution
  are compared across input count rates and three filter types ($\tau=6$
  ms), for both short (10.24 ms) and long (25.60 ms) pulse records.  We note
  that the maximum output pulse rate is considerably lower than the
  corresponding raw pulse rate, because many raw pulses are due to spectral
  features other than the 97.431 keV line.}
\end{figure}

In Fig.~\ref{icr_ocr} the output pulse rate, for the energy range
$97.431\pm 0.100$ keV, and energy resolution are compared for all four
input count rates and the three types of filter, for both short and
long pulse records.  At the highest rate and for short pulse records,
the filter orthogonal to both constants and exponentials ($\tau=6$ ms)
offers 45 \% higher output rate than the standard DFT-computed filter
and 40 \% higher than the filter orthogonal to constants alone, at
better energy resolution than either one.

One important issue regarding the filters orthogonal to exponentials
concerns their performance sensitivity to the choice of decay time constant
$\tau$.  The average pulse, for the TES microcalorimeter tested, was
well-approximated over a 25.6 ms record by a linear combination of
four exponentials, with decay time constants $\tau=$ 0.018 ms, 0.144 ms, 0.963
ms, and 2.514 ms.  If just the tail is fit, however, the constants
increase considerably.  It is evident, therefore, that no single decay
rate is optimal for all arrival-time lags.  Nevertheless, for the full
set of Poisson-distributed arrival time lags, the performance of the
filters is only mildly sensitive to the choice of time constant.  For
the highest-rate data with short records, over the range
$\tau=3,\ldots,10$ ms, the $97.431\pm0.100$ keV output pulse rate
varied as $1.568\pm 0.044$ Hz (mean and one standard deviation) and
the energy resolution varied as $128.3\pm 3.2$ eV, as compared with
the $\tau=6$ ms values of 1.640 Hz and 125.7 eV.

\paragraph*{Summary.}
The proposed filter construction method, differing from the
standard procedure by being computed in the time domain and enabling
filter optimization subject to explicit length and orthogonality
constraints, assumes linear superposition of pulses and simple
exponential decay of pulse tails.  Although these assumptions are
satisfied rather imperfectly for the TES microcalorimeter tested, the
method yields notable improvement over standard filtering.  Our tests
also point to additional, more specialized options, such as filters
optimized for a particular interval of arrival time lags since the
previous pulse.  Such filters fit easily within this framework and
underline that the new approach has implications for pulse processing
in a broad range of applications.

We gratefully acknowledge support from the NIST Innovations in Measurement
Science program, the DOE Office of Nuclear Nonproliferation Research and
Development, and the DOE Office of Nuclear Energy.

\bibliography{filters}

\end{document}